\documentclass[useAMS,usenatbib]{mn2e}

\usepackage{multirow}
\usepackage{booktabs}
\usepackage{epsfig}
\usepackage{color}
\usepackage{ulem}
\def\arcsec{\hbox{$^{\prime\prime}$}}

\def\lesssim{\mathrel{\hbox{\rlap{\hbox{\lower3pt\hbox{$\sim$}}}\hbox{\raise1pt\hbox{$<$}}}}}
\def\gtrsim{\mathrel{\hbox{\rlap{\hbox{\lower3pt\hbox{$\sim$}}}\hbox{\raise1pt\hbox{$>$}}}}}

\newcommand\aj{\rmfamily{AJ}}% 
\newcommand\apj{\rmfamily{ApJ}}% 
\newcommand\apjl{\rmfamily{ApJ}}% 
\newcommand\apjs{\rmfamily{ApJS}}% 
\newcommand\aap{\rmfamily{A\&A}}% 
\newcommand\mnras{\rmfamily{MNRAS}}% 
\newcommand\pasa{\rmfamily{PASA}}% 
\newcommand\pasp{\rmfamily{PASP}}% 
\makeatletter

\newcommand{\Rmnum}[1]{\expandafter\@slowromancap\romannumeral #1@}
\makeatother

\newcommand\RSAA{Research School of Astronomy and Astrophysics, The Australian National University, Canberra, ACT 2611, Australia}
\newcommand\CAASTRO{ARC Centre of Excellence for All-sky Astrophysics (CAASTRO)}
\newcommand\AAO{Australian Astronomical Observatory, North Ryde, NSW 2113, Australia}
\newcommand\UQ{School of Mathematics and Physics, University of Queensland, QLD 4072, Australia}
\newcommand\Swinburne{Centre for Astrophysics and Supercomputing, Swinburne University of Technology, Hawthorn, VIC 3122, Australia}
\newcommand\USyd{Sydney Institute for Astronomy, School of Physics, A28, The University of Sydney, NSW, 2006, Australia}
\newcommand\CSIRO{CSIRO Astronomy \& Space Science, Epping, NSW 1710, Australia}
\newcommand\Dark{Dark Cosmology Centre, Niels Bohr Institute, University of Copenhagen, Juliane Maries Vej 30, DK-2100 Copenhagen, Denmark}

\newcommand\NDphy{Department of Physics and JINA Center for the Evolution of the Elements, University of Notre Dame, 225 Nieuwland Science       Hall, Notre Dame, IN 46556, USA}
\newcommand\INAF{INAF- Astrophysics Observatory of Turin , Strada Osservatorio 20, Pino Torinese, Torino, 10025, Italy}
\newcommand\Southampton{School of Physics and Astronomy, University of Southampton, Southampton, SO17 1BJ, UK}

\newcommand\ANL{Argonne National Laboratory, 9700 South Cass Avenue, Lemont, IL 60439, USA}
\newcommand\BNL{Brookhaven National Laboratory, Bldg 510, Upton, NY 11973, USA}
\newcommand\OSUccapp{Center for Cosmology and Astro-Particle Physics, The Ohio State University, Columbus, OH 43210, USA}
\newcommand\CIEMAT{Centro de Investigaciones Energ\'eticas, Medioambientales y Tecnol\'ogicas (CIEMAT), Avda. Complutense 40, Madrid, Spain}
\newcommand\CTIO{Cerro Tololo Inter-American Observatory, National Optical Astronomy Observatory, Casilla 603, La Serena, Chile}
\newcommand\OSUastro{Department of Astronomy, The Ohio State University, Columbus, OH 43210, USA}
\newcommand\UIUCastro{Department of Astronomy, University of Illinois at Urbana-Champaign, 1002 W.\ Green St., Urbana, IL 61801, USA}
\newcommand\UMastro{Department of Astronomy, University of Michigan, Ann Arbor, MI 48109, USA}
\newcommand\UCL{Department of Physics \& Astronomy, University College London, Gower Street, London WC1E 6BT, UK}
\newcommand\Sussex{Astronomy Centre, University of Sussex, Falmer, Brighton, BN1 9QH, UK}
\newcommand\UP{Department of Physics and Astronomy, University of Pennsylvania, Philadelphia, PA 19104, USA}
\newcommand\LMU{Department of Physics, Ludwig-Maximilians-Universit\"{a}t, Scheinerstr.\ 1, 81679, M\"{u}nchen, Germany}
\newcommand\Stanfordphy{Department of Physics, Stanford University, 382 Via Pueblo Mall, Stanford, CA 94305, USA}
\newcommand\OSUphy{Department of Physics, The Ohio State University, Columbus, OH 43210, USA}
\newcommand\UIUCphy{Department of Physics, University of Illinois at Urbana-Champaign, 1110 W.\ Green St., Urbana, IL 61801, USA}
\newcommand\UMphy{Department of Physics, University of Michigan, Ann Arbor, Michigan 48109, USA}
\newcommand\FNAL{Fermi National Accelerator Laboratory, P.O. Box 500, Batavia, IL 60510 USA}
\newcommand\TAMU{George P. and Cynthia Woods Mitchell Institute for Fundamental Physics and Astronomy, and Department of Physics and Astronomy, Texas A \& M University, College Station, TX 77843-4242, USA}
\newcommand\ICRA{ICRA, Centro Brasileiro de Pesquisas F\'isicas, Rua Dr. Xavier Sigaud 150, CEP 22290-180, Rio de Janeiro, RJ, Brazil}
\newcommand\ICREA{Instituci\'o Catalana de Recerca i Estudis Avan\c{c}ats, E-08010 Barcelona, Spain}
\newcommand\IAP{Institut d'Astrophysique de Paris, Univ. Pierre et Marie Curie \& CNRS UMR7095, F-75014 Paris, France}
\newcommand\IEEC{Institut de Ci\`encies de l'Espai, IEEC-CSIC, Campus UAB, Facultat de Ci\`encies, Torre C5 par-2, 08193 Bellaterra, Barcelona, Spain}
\newcommand\IFAE{Institut de F\'{\i}sica d'Altes Energies, Universitat Aut\`{o}noma de Barcelona, E-08193 Bellaterra, Barcelona, Spain}
\newcommand\Portsmouth{Institute of Cosmology and Gravitation, University of Portsmouth, Dennis Sciama Building, Burnaby Road, Portsmouth, PO1 3FX, UK}
\newcommand\JPL{Jet Propulsion Laboratory, California Institute of Technology, 4800 Oak Grove Dr., Pasadena, CA 91109, USA}
\newcommand\KavliChicago{Kavli Institute for Cosmological Physics, University of Chicago, Chicago, IL 60637, USA}
\newcommand\KavliCam{Kavli Institute for Cosmology, University of Cambridge, Madingley Road, Cambridge CB3 0HA, UK}
\newcommand\KavliStanford{Kavli Institute for Particle Astrophysics \& Cosmology, P. O. Box 2450, Stanford University, Stanford, CA 94305, USA}
\newcommand\LIneA{Laborat\'orio Interinstitucional de e-Astronomia - LIneA, Rua Gal. Jos\'e Cristino 77, Rio de Janeiro, RJ - 20921-400, Brazil}
\newcommand\LBNL{Lawrence Berkeley National Laboratory, 1 Cyclotron Road, Berkeley, CA 94720, USA}
\newcommand\MPI{Max-Planck-Institut f\"{u}r extraterrestrische Physik, Giessenbachstr.\ 85748 Garching, Germany}
\newcommand\NCSA{National Center for Supercomputing Applications, 1205 West Clark St., Urbana, IL 61801, USA}
\newcommand\Cambridge{Institute of Astronomy, University of Cambridge, Madingley Road, Cambridge CB3 0HA, UK}
\newcommand\ON{Observat\'orio Nacional, Rua Gal. Jos\'e Cristino 77, Rio de Janeiro, RJ - 20921-400, Brazil}
\newcommand\SLAC{SLAC National Accelerator Laboratory, Menlo Park, CA 94025, USA}
\newcommand\Munich{University Observatory Munich, Scheinerstrasse 1, 81679 Munich, Germany}

\newcommand\UArizona{Department of Physics, University of Arizona, 1118 E 4th St, Tucson, AZ 85721, USA}

\newcommand\CAPES{CAPES Foundation, Ministry of Education of Brazil, Bras\'{i}lia - DF 70040-020, Brazil}

\title[OzDES Y1]{OzDES multi-fibre spectroscopy for the Dark Energy Survey: first-year operation and results}
\author[F.~Yuan et al]
{\parbox{\textwidth}{
Fang~Yuan,$^{1,2}$\thanks{E-mail: fang.yuan@anu.edu.au}
C.~Lidman$^{2,3}$,
T.~M.~Davis$^{2,4}$,
M.~Childress$^{1,2}$,
F.~B.~Abdalla$^{5}$,
M.~Banerji$^{6,7}$,
E.~Buckley-Geer$^{8}$,
A.~Carnero~Rosell$^{9,10}$,
D.~Carollo$^{11,12}$,
F.~J.~Castander$^{13}$,
C.~B.~D'Andrea$^{14}$
% ...
}\vspace{0.4cm}\\
Affiliations are listed at the end of the paper
}
\begin{document}

\date{Accepted -  Received -}

\pagerange{\pageref{firstpage}--\pageref{lastpage}} \pubyear{????}

\maketitle

\label{firstpage}

\clearpage

\begin{abstract}

OzDES is a five-year, 100-night, spectroscopic survey on the
Anglo-Australian Telescope, whose primary aim is to measure redshifts of
approximately 2,500 Type Ia supernovae host galaxies over the redshift range $0.1 < z < 1.2$, and derive
reverberation-mapped black hole masses for approximately 500
active galactic nuclei and quasars over $0.3 < z < 4.5$.  This treasure
trove of data forms a major part of the spectroscopic follow-up for
the Dark Energy Survey for which we are also targeting cluster
galaxies, radio galaxies, strong lenses, and unidentified transients,
as well as measuring luminous red galaxies and emission line
galaxies to help calibrate photometric redshifts.

Here we present an overview of the OzDES program and our first-year
results. Between Dec 2012 and Dec 2013, we observed over 10,000 objects and
measured more than 6,000 redshifts.  Our strategy of retargeting faint
objects across many observing runs has allowed us to measure redshifts
for galaxies as faint as $m_r=25$ mag.  We outline our target selection and
observing strategy, quantify the redshift success rate for different
types of targets, and discuss the implications for our main science
goals.  Finally, we highlight a few interesting objects as examples of the fortuitous yet not
totally unexpected discoveries that can come from such a large
spectroscopic survey.

\end{abstract}


\section{INTRODUCTION}
\label{sec:intro}

The Australian Dark Energy Survey (OzDES)\footnote{Or: {\it Optical
    redshifts for the Dark Energy Survey}.} has been designed to
provide efficient spectroscopic follow-up of targets identified from imaging by the Dark
Energy Survey \citep[DES;][]{2005IJMPA..20.3121F,2014SPIE.9149E..0VD}.  OzDES extends DES by enabling
new science goals that cannot be achieved without spectroscopic
information --- such as supernova cosmology and reverberation mapping
of active galactic nuclei (AGN).  We also enhance DES by providing an
important source of calibration data for photometric redshifts, which
are the cornerstone for the majority of the DES science programs.

\begin{figure}
\includegraphics[width=84mm]{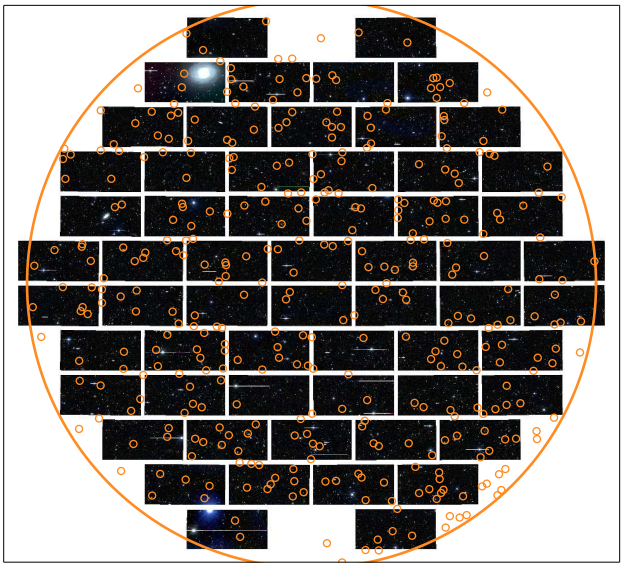}
\caption{The 2dF spectrograph has 392 optical fibres to be placed at
  positions across a field of view 2.1 degree
  in diameter (large orange circle).  This is well matched to the DECam field
  of view (image mosaic in the background), making
  spectroscopic follow-up very efficient. The small orange circles represent
  the locations of targets selected for one set of AAT observations.}
\label{fig:fov}
\end{figure}

The 3.9m Anglo-Australian Telescope (AAT) used by OzDES and the CTIO~4m
Blanco telescope used by DES are ideal partners for spectroscopy and
imaging because they have well matched $\sim$2 degree diameter fields of view (see
Fig.~\ref{fig:fov}).  The AAT and CTIO~4m are similar in several
respects --- both are 4m class telescopes that were commissioned in
1974 and both have recently been rejuvenated with powerful new
instrumentation: in the case of CTIO it is the 570 mega-pixel Dark
Energy Camera \citep[DECam;][]{2012PhPro..37.1332D,2012SPIE.8446E..11F}, while on the AAT it is the new efficient
AAOmega spectrograph \citep{2004SPIE.5492..410S} coupled with the Two
Degree Field (2dF) 400-fibre multi-object fibre-positioning system
\citep{2002MNRAS.333..279L}.

The DES program consists of a wide-field survey covering 5000 square
degrees, as well as a rolling survey of ten fields that cover a total of 30
square degrees \citep{2014SPIE.9149E..0VD}.  These ten fields (see Table~\ref{tab:field_centers} for the sky coordinates) are targeted repeatedly over the
course of the survey with a cadence of approximately 6 days in order
to find transient objects, such as supernovae, and monitor variable
objects, such as AGN.  OzDES repeatedly targets these ten fields,
selecting objects that range in brightness from $m_r\sim17$ to 
$m_r\sim25$ mag, a range of more than a thousand in flux density.  Once a
redshift is obtained, we deselect the target (with some exceptions).  
Objects that lack a redshift are
observed until a redshift is measured. 
This tactic allows us to obtain
redshifts for targets far fainter than ever previously achieved with
the AAT.  Together this means we can run an efficient survey of bright
objects while simultaneously acquiring spectra for much fainter
objects.

\begin{table}
\caption{Center coordinates of the ten DES SN fields.}
\label{tab:field_centers}
\begin{tabular*}{80mm}{@{\extracolsep{\fill}}lccc}
\hline
Field Name & R.A. (h m s) & Decl. ($\circ$ $^\prime$ ${''}$) & Comment\textsuperscript{*} \\
\hline
~E1 & 00 31 29.9 & -43 00 35 & \multirow{2}{*}{ELAIS\textsuperscript{$\dagger$} S1} \\
~E2 & 00 38 00.0 & -43 59 53 & \\ 
\hline
~S1 & 02 51 16.8 & ~00 00 00 & \multirow{2}{*}{Stripe 82\textsuperscript{$\ddagger$}} \\
~S2 & 02 44 46.7 & -00 59 18 & \\
\hline
~C1 & 03 37 05.8 & -27 06 42 & \multirow{3}{*}{CDFS\textsuperscript{$\S$}} \\
~C2 & 03 37 05.8 & -29 05 18 & \\
~C3 (deep) & 03 30 35.6 & -28 06 00 & \\
\hline
~X1 & 02 17 54.2 & -04 55 46 & \multirow{3}{*}{XMM-LSS\textsuperscript{$\parallel$}} \\
~X2 & 02 22 39.5 & -06 24 44 & \\
~X3 (deep) & 02 25 48.0 & -04 36 00 & \\
\hline
\\
\multicolumn{4}{l}{\textsuperscript{*}\footnotesize{Overlap with areas covered by other surveys.}} \\
\multicolumn{4}{l}{\textsuperscript{$\dagger$}\footnotesize{European Large-Area ISO Survey.}} \\
\multicolumn{4}{l}{\textsuperscript{$\ddagger$}\footnotesize{An equatorial region repeatedly imaged by SDSS.}} \\
\multicolumn{4}{l}{\textsuperscript{$\S$}\footnotesize{Chandra Deep Field South survey.}} \\
\multicolumn{4}{l}{\textsuperscript{$\parallel$}\footnotesize{X-ray Multi Mirror Large Scale Structure survey.}} \\

\end{tabular*}
\end{table}

OzDES has a total of 100 nights distributed across five
years during the DES observing seasons (August to January).  The first
season of OzDES began in 2013B (August 2013 to January 2014) with 12 nights (designated as Y1).
Allocations will progressively increase each year, to
accommodate the increasing number of supernovae host galaxies that DES will 
have accumulated in subsequent years. 

In the 2012B (August 2012 to January 2013) semester, a year before the start of DES and OzDES, the
DECam was used to execute the DES SN program as part of
its Science Verification (SV) phase of commissioning.  In parallel,
AAT/AAOmega-2dF time was awarded for the observation of the 
DES SN fields in a precursor program to OzDES. 
Supplementary observing time (1 night during SV and 2 nights during Y1) was also obtained through the National Optical Astronomy Observatory (NOAO).

In this paper, we present an overview of OzDES and the 
results obtained during the SV and Y1 seasons. 
Non-DES fields and targets observed in the SV season are excluded from the analysis.
When appropriate, we also discuss changes and improvements to the OzDES in Y2 (2014B, August 2014 to January 2015).

We organise the paper as follows: In $\S$~\ref{sec:science} we
summarise our science goals and in $\S$~\ref{sec:strategy} we
present the operational details that were current at the end of Y1,
highlighting what was modified to improve the efficiency of the
survey for Y2. Results of redshifts and quality assessments are
presented in $\S$~\ref{sec:results}, followed by discussions of
implications for science in $\S$~\ref{sec:discussion}. Finally, we
conclude in $\S$~\ref{sec:conclusion}.

\section{Science Goals}\label{sec:science}

Here we outline the wide range of science goals we aim to achieve with
OzDES. Each science goal may require one or more types of target and
the targeting strategies evolve with time. Details of the target types used 
in SV and Y1 are described in $\S$~\ref{sec:targets}.

\subsection{Type Ia Supernova Cosmology}

The main science motivation of OzDES is to obtain host galaxy spectroscopic
redshifts for a sample of 2,500 Type Ia SNe discovered by DES, 
with the goal of improving measurements of the Universe's
 global expansion history.
This goal is efficiently achieved with AAT's unique combination of
multi-object fibre-fed spectroscopy and wide field of view.

Spectroscopy of live supernovae within a few weeks of maximum light
serves the dual purpose of typing the transient (determining whether
it is a Type Ia SN) and measuring its redshift.  However, the spatial
density of SNe Ia that are bright enough to be observed with the AAT
at any instant in time is only a handful per square degree. An 8--10 metre
class telescope is often required for such observations, but can
usually only observe one object at a time. This mode of follow-up is
thus time-consuming and requires an unrealistically large quantity of
resources to acquire a large sample.

Alternatively, well sampled multicolour light curves can be used to
reliably identify a SN Ia \citep{2011ApJ...738..162S}, after which a
spectrum of the host galaxy can be used to measure its precise redshift.  
Host galaxy redshifts can also be used to improve photometric
classification \citep{2014AJ....147...75O}.

Host galaxies can be observed at any time, even after the supernova has
faded. Thus one can wait to collect many supernovae in a field before
measuring their host-galaxy redshifts efficiently with multi-object
spectroscopy.  This strategy has been tested in the Sloan Digital Sky
Survey \citep[SDSS-II,][]{2014AJ....147...75O} and the SuperNova Legacy
Survey \citep[SNLS,][]{2013PASA...30....1L}.

\subsection{AGN Reverberation Mapping}

The other primary science goal of OzDES is reverberation mapping of
AGN and quasars. Reverberation mapping
\citep{1982ApJ...255..419B,1993PASP..105..247P} is an effective way to
measure supermassive black hole masses in AGN, 
over much of the age of the
universe. This is possible because the continuum emission from the AGN
accretion disk is variable, and this continuum emission photoionizes
the clouds of gas at larger scales that give rise to the
characteristic broad emission lines of most AGN. As the continuum
emission varies in intensity, the broad emission lines reverberate in
response with a time delay that depends on the light travel time from
the continuum source.

Measurement of this time delay provides a geometric size for the broad
line region. This size scale
can be used to measure the mass of the supermassive black hole
through application of the virial theorem and measurement of the
velocity width of the broad emission line. Approximately 50 AGN have
reverberation-based black hole masses to date, and the masses of these
black holes agree well with mass measurements from stellar dynamics
\citep{2006ApJ...646..754D, 2014ApJ...791...37O} and yield the same
slope as the $M-\sigma_*$ relation that holds for quiescent galaxies
\citep{2010ApJ...716..269W, 2013ApJ...773...90G}.
The size scale, typically determined from the H$\beta$
emission line, is also very well correlated with the AGN luminosity with
the $R \propto L^{1/2}$ scaling expected from a simple photoionization
model \citep[e.g.][]{2013ApJ...767..149B}. The relatively small
scatter in this relation was used by \citet{2011ApJ...740L..49W} to
demonstrate that AGN could be used as standard candles. 

The current sample of about 50 AGN with reverberation-based masses are
all in low to moderate luminosity AGN, and nearly all in the
relatively nearby universe ($z<0.3$). This is because reverberation
mapping requires a substantial amount of telescope time to measure the
time lags, and it has proven most straightforward to get the necessary
allocation to observe bright objects with small telescopes. The
lower-luminosity AGN also have lags of only days to weeks, and thus
can be measured with a single semester of data.  It is much more
difficult to measure the year or longer lags of the most luminous AGN
at redshifts $z>1$ and higher \citep[although
  see][]{2007ApJ...659..997K}; yet these AGN are arguably the most
interesting as they represent the majority of supermassive black hole
growth in the universe.  OzDES is presently monitoring $\sim$1000 AGN
up to $z \approx 4$ and aims to measure reverberation lags and black
hole masses for approximately 40\% of the final sample (King et
al., in prep). This new, multi-object reverberation mapping project, as
well as other similar efforts \citep{2015ApJS..216....4S}, will
provide a wealth of new data on black hole masses out to and beyond
the peak of the AGN era. We will also use new measurements of the
radius-luminosity relation to construct a Hubble diagram out to higher
redshifts than can be reached with supernovae, which provides some
complementary constraints on the time variation in dark energy
\citep{2014MNRAS.441.3454K}.

\subsection{Transients}

Concurrent OzDES and DES observing enables time-critical spectroscopic 
observations of transients discovered in imaging.
With a monthly observing cadence, OzDES is expected to target
several hundred active transients, putting these at highest priority.  
This sample, supplemented by
observations of fainter events by larger telescopes, will provide
crucial validation of the photometric SN classification 
and enable detailed studies of these SNe.

We target all kinds of transient candidates,
including those with uncertain physical nature. A survey of this size and
scope expects to find surprises in the data.  With our targeting
strategy we aim to investigate the unexpected and potentially find
as-yet unidentified classes of transients.

\subsection{Photo-$z$ training}

A core requirement of DES is to obtain accurate photometric redshifts
(photo-$z$) for the majority of galaxies in the wide survey.  This will
enable key science goals, such as the measurement of baryon acoustic
oscillations (BAO) with millions of galaxies, and the use of weak
lensing for cosmology.  Our spectra play an important role in
providing a spectroscopic sample for calibrating and testing the DES
photometric redshifts, and a significant number of our fibres are
allocated to Luminous Red Galaxies (LRG), Emission Line Galaxies
(ELG), and other photo-$z$ targets.

Redshifts from OzDES have already been used in a number of recently
published studies on DES photometric
redshifts. \citet{2014MNRAS.445.1482S} have used the redshifts to
evaluate the performance of various photo-$z$ methods on DES SV data
and found several codes to produce photo-$z$ precisions and outlier
fractions that satisfy DES science requirements.
\citet{2015MNRAS.446.2523B} combines optical data from DES and
near-infrared (NIR) data from VISTA Hemisphere Survey \citep[VHS;][]{2013Msngr.154...35M}
 to improve photo-$z$ performance.  In particular, selection
criteria based on optical-NIR colors are applied to identify LRG
targets at high redshift ($z\gtrsim0.5$) for OzDES. Spectroscopic
results are used to verify the effectiveness of this selection method.

\subsection{Radio galaxies, cluster galaxies, and strong lenses}

The large number of fibres available to 2dF allows us to pursue
a wide range of supplementary science goals.  These include the
following:

\begin{itemize}

\item {Gathering redshifts of galaxies selected from the ATLAS \citep[Australia Telescope Large Area Survey;][Franzen et al. in prep, Banfield et al. in prep]{2006AJ....132.2409N}. ATLAS is a deep, 14/17 uJy/beam rms, 1.4 GHz survey of 3.6/2.7 deg$^2$ of the CDFS/ELAIS-S1 survey fields which have \textgreater90\% overlap with the DES deep imaging. ATLAS is being used to study the astrophysics of radio sources, and is also being used as a pathfinder to develop the science and techniques for the  primary radio continuum survey \citep[EMU;][]{2011PASA...28..215N} of the Australian Square Kilometre Array Pathfinder (ASKAP). ATLAS has detected over 5000 radio sources of which at least half will be targeted by OzDES over the course of the survey. These redshifts will be used to determine the evolution of the faint radio population, including both star forming galaxies and radio AGN, up to redshift greater than one. They will also be used to calibrate photometric and statistical redshift algorithms for use with the 70 million EMU sources (for which spectroscopy is impractical). Furthermore, the detected optical emission lines will provide insights into the detailed astrophysics within these galaxies, including distinguishing star forming galaxies from AGN.}

\item {Confirming previously unknown cluster candidates and
gathering redshifts for cluster galaxies, especially 
central galaxies for the calibration of the cluster red sequence, as 
well as validation of cluster photometric redshifts. 
Our repeated returns to the same field allow us to collect
redshifts for multiple galaxies in a single cluster. Usually this is 
impossible for all but very nearby clusters because the instrumental 
limit of fibre collisions prevents one from measuring closely 
neighbouring galaxies in the same exposure.}

\item {Measuring redshifts for both the lens and
the background lensed galaxy or quasar in strong lens candidates.
Some of the lensed quasar targets may be suitable for time-delay experiments.} 

\end{itemize}

\subsection{Calibration}

About 10\% of fibres are used for targets that facilitate the
the calibration of the data. These include:

\begin{itemize}

\item Regions that are free of objects (sky fibres). Some of our
  targets are 100 times or more fainter than the sky in the $2{\arcsec}$ fibre aperture, so a good
  estimate of the sky brightness is crucial.

\item F stars that are used to monitor throughput, which is heavily
  dependent on the seeing and the amount of cloud cover. Up to 15
  F stars are observed per field and used to derive a mean sensitivity
  curve. The variation of the sensitivity curve over each plate allows
  for an estimate of the accuracy of the flux calibration and the
  mean value allows us to appropriately weight data that are
  obtained over multiple occasions.

\item Candidate hot (T$_{\rm eff} \sim 20,000~\rm{K}$) DA (hydrogen
  atmosphere) white dwarfs (WDs) that can be used as primary flux
  calibrators for the DES deep fields. Stellar atmosphere modeling
  uncertainties for hot DA WDs are small so that synthetic
  photometry can be compared with DES observations with an expected
  accuracy of 2--3\% or better per candidate. A collection of $\sim$100
  such candidates over the DES deep fields will allow one to test the
  accuracy of the photometric calibration of the DES deep fields. The
  number of known DA WDs in the DES deep fields is currently
  too small to make this test, so we aim to find new ones.

\end{itemize}

\section{Observing Strategy}\label{sec:strategy}

\subsection{Instrument Setup and Observations}\label{sec:setup}

The 2dF robotic positioner allows up to 392 targets to be observed
simultaneously over a field of view 2.1 degrees in diameter (there are also 8 fibre bundles for guiding). 
The projected
fibre diameter is approximately two arcsec. Two sets of fibres are
provided on separate field plates mounted back-to-back on a
tumbler. Configuration of all fibres on a single plate takes about
40 minutes and can be done as the other plate is being observed,
thereby greatly reducing overheads. A minimum separation of 30 to 40
arcsec between fibres is imposed by the physical size of the
rectangular fibre buttons. This constraint and other hardware limits
are respected by the custom fibre configuration software.

The fibres feed AAOmega, which is a bench mounted double beam
spectrograph sitting in one of the Coude rooms of the AAT. The light from
the fibres is first collimated with a mirror before passing through a
dichroic which splits the light at 570~nm into two arms, one red and
one blue. In the blue arm, we used the 580V grating (dispersion of 1~A per pixel). In the red arm,
we used the 385R grating (dispersion of 1.6~A per pixel). The resulting wavelength coverage starts at
370~nm and ends at 880~nm, with a resolution of R$\sim$1400. 
Up until the beginning of 2014, the
detectors were two 2k$\times$4k E2V CCDs. These detectors
were replaced with new, cosmetically superior and more efficient 2k$\times$4k E2V CCDs during 2014 
\citep{Brough2014}. 

The full SV and Y1 observing log is shown in
Table~\ref{tab:observing_log}. The standard observing sequence for a
field configuration is 3 consecutive 40 minute exposures. In practice,
the sequence is constrained by observing conditions and field
observability.  Among the ten DES SN fields, the two deep fields (C3
and X3) take priority in the scheduling and have accumulated the
longest integration time, about 16 hours compared to an average of 7
hours for the shallow ones.

For each configuration, we took a single arc and up to two fibre
flats. The arc is used for wavelength calibration, while the flats are
used to define the location of the fibres on the detector (the
so-called tram line map) and to determine the relative chromatic
throughput of the fibres.

At the end of each run the nightly bias and dark frames are combined
to produce a master bias frame and a master dark frame. 
During our first observing season,
the blue CCD contained several notable defects. The
master bias and master dark were used to mitigate the impact of these
defects on the science exposures. By comparison, the red CCD was 
cosmetically superior, so the corresponding
master bias and master dark for the red CCD were not needed (indeed,
applying the correction just added noise). 

\begin{table*}
\centering
\begin{minipage}{\textwidth}
\caption{OzDES first year observing log for DES SN fields.}
\label{tab:observing_log}
\begin{tabular*}{\textwidth}{@{\extracolsep{\fill}}ccccccccccccc}
\hline
\multirow{2}{*}{UT Date} & \multirow{2}{*}{\parbox{1.5cm}{\centering Observing Run}} & \multicolumn{10}{c}{Total exposure time for DES field (minutes)} & \multirow{2}{*}{Note} \\
\cmidrule{3-12} 
& & E1 & E2 & S1 & S2 & C1 & C2 & C3(deep) & X1 & X2 & X3(deep)\\
\hline
2012-12-13 &  \multirow{4}{*}{001} & -- 	& -- 	& -- 	& -- 	& 40	& 80 	& -- 	&120& -- 	& --  & \\
2012-12-14 &         & -- 	& -- 	& -- 	& -- 	&120& 80 & -- 	& -- 	& -- 	& --   \\
2012-12-15 &         & -- 	& -- 	& -- 	& -- 	& -- 	& 40 &120& -- 	& -- 	& --   \\
2012-12-16 &         & -- 	& -- 	& -- 	& -- 	& -- 	& -- 	& 80 & -- 	& -- 	&160\\
\hline
2013-01-05\textsuperscript{*} & 002 & -- 	& -- 	& -- 	& -- 	& -- 	& -- 	&120& -- 	& -- 	& 80 & \\
\hline
2013-09-29\textsuperscript{$\dagger$} &  \multirow{3}{*}{003} & -- 	& -- 	& -- 	& -- 	& -- 	& -- 	& 40 	& -- 	& -- 	& --  & \multirow{3}{*}{\parbox{2.8cm}{About one night lost due to bad weather.}} \\
2013-09-30 &         & -- 	&120& -- 	& -- 	& -- 	&100& 80 & -- 	& -- 	&120\\
2013-10-01 &         & -- 	& -- 	& -- 	& -- 	&150& -- 	& -- 	& -- 	& -- 	& --   \\
\hline
2013-10-30 &  \multirow{6}{*}{004} &120 	& -- 	& -- 	& -- 	& 80 	& -- 	&120& -- 	& -- 	&120& \multirow{6}{*}{\parbox{3cm}{About one night lost due to hardware problem.}}\\
2013-10-31 &         & -- 	&120& -- 	& -- 	& -- 	& -- 	& -- 	& -- 	& -- 	& --   \\
2013-11-01 &         &120 	& -- 	&120& -- 	& -- 	& 80 	& -- 	&120& -- 	& --   \\
2013-11-02 &         & -- 	& -- 	& -- 	&120& 80 & -- 	& -- 	& -- 	& 40 	& --   \\
2013-11-03 &         & -- 	&120& -- 	& -- 	& -- 	& -- 	&120& -- 	& 80 &120\\
2013-11-04 &         &120 	& -- 	&120& -- 	& -- 	& 80 	& -- 	&120& -- 	& --   \\
\hline
2013-11-28 &  \multirow{4}{*}{005} & -- 	& -- 	& -- 	& -- 	& -- 	& -- 	& 14	& -- 	& -- 	&120& \multirow{4}{*}{\parbox{3cm}{About one night lost due to bad weather.}} \\
2013-11-29 &         & -- 	& -- 	& -- 	& -- 	& -- 	& -- 	&120& -- 	& -- 	&120\\
2013-11-30 &         &120 	& -- 	& -- 	& -- 	& 40 	&120& -- 	& -- 	&120& --   \\
2013-12-01 &         & -- 	&120& 40 & -- 	&107& -- 	& -- 	&120& -- 	& --   \\
\hline
2013-12-25\textsuperscript{*} &  \multirow{2}{*}{006} & -- 	& -- 	& -- 	& -- 	& -- 	& -- 	& -- 	& -- 	& -- 	& --  & \multirow{2}{*}{\parbox{3cm}{About one night lost due to bad weather.}} \\
2013-12-26\textsuperscript{*} &         & -- 	& -- 	& -- 	& -- 	& 40 	& -- 	&120& -- 	& -- 	&120\\
\hline
{\bf 2012--2013} & {\bf Total (mins)} & 480 & 480 & 280 &120& 657 & 580 & 934 & 480 & 240 & 960 \\
\hline
\\

\multicolumn{13}{l}{\textsuperscript{*}\footnotesize{NOAO time allocation.}}\\
\multicolumn{13}{l}{\textsuperscript{$\dagger$}\footnotesize{C3 field was observed at the end of the night during time allocated to the XMM-XXL collaboration.}} \\
\end{tabular*}
\end{minipage}
\end{table*}

\subsection{Target Selection}

There are three stages to target selection: 

\begin{description}
\item[i)] creation of the input catalogues, 
\item[ii)] target prioritisation, and 
\item[iii)] target allocation.  
\end{description}

For each field, the input catalogues contain a large number of potential targets (step
i), from which we select a prioritised set of 800 potential targets
(step ii), which the 2dF configuration software uses to optimally
allocate its fibres for each observation (step iii).
 
In the first stage, science targets are provided by the science
working groups within DES and OzDES. The working groups are
responsible for updating their input catalogs between observing runs,
e.g. removing objects that have reliable redshifts from earlier runs.

The number of targets greatly exceeds the number of fibres, so not all
targets can be observed. Based on the scientific importance of the
targets, we assign a priority (larger number for higher priority) and
a quota to each type of object, as defined in
Table~\ref{tab:target_type}.

From the initial input catalog, the prioritised target list is
selected based on priority and quota, starting with the highest
priority and ending at priority 4. 
Targets of a particular type are randomly selected up to its quota.
If an object cannot be selected
because the quota has been reached,
then the object goes into one of two pools. Objects with priorities
six and above go into a high priority backup pool (priority 3).  All
other objects, including objects for photo-$z$ calibration go into a
low priority pool (priority 2). If at the end of this initial
allocation the number of objects is less than 800, then objects from
the high priority backup pool are randomly selected until 800 objects have been
chosen. If the total number of objects is still less than 800, then
objects from the low priority backup pool are selected until 800
objects have been chosen. 

In the third stage, targets are allocated to fibres using the custom
2dF fibre configuration software.  This software avoids fibre
collisions (see $\S$~\ref{sec:setup}) while optimising the target
distribution so the highest-priority targets are preferentially
observed. A maximum of 392 fibres are allocated in this step including
25 for sky positions.  Another 8 fibres are placed on bright sources
for guiding. 
The input list size of 800 balances efficiency and performance 
of the fibre configuration software. 
A sufficient number of objects is required for a good spacial distribution but
the algorithm slows down significantly for a larger number of targets than 800.

If a field is observed a second time during an observing run, the
above three-step allocation process is repeated.  Usually the only
change to the input catalogues is the inclusion of just-discovered
transients.  During target prioritisation, we deselect targets that
have been given secure redshifts from the previous nights, freeing
those fibres for new targets.  The priority of objects that have been
observed in the current run, but do not satisfy the deselection
criteria (defined in Table~\ref{tab:target_type}) are boosted by an
amount that depends on their initial priority.

\subsection{Target Definition}\label{sec:targets}

Here, we define the target types and their related targeting strategy used during SV and Y1, ordered by
object priority from highest to lowest. If more than one priority is available for a type, its location in the list is determined by the higher priority. 
In general, we do not expect the
list to evolve significantly, especially for targets that have high
priority. However, it is likely that we will modify the quotas as the
survey progresses so as to maximise the results of the survey.

\begin{figure*}
\begin{minipage}{\textwidth}
\includegraphics[width=180mm]{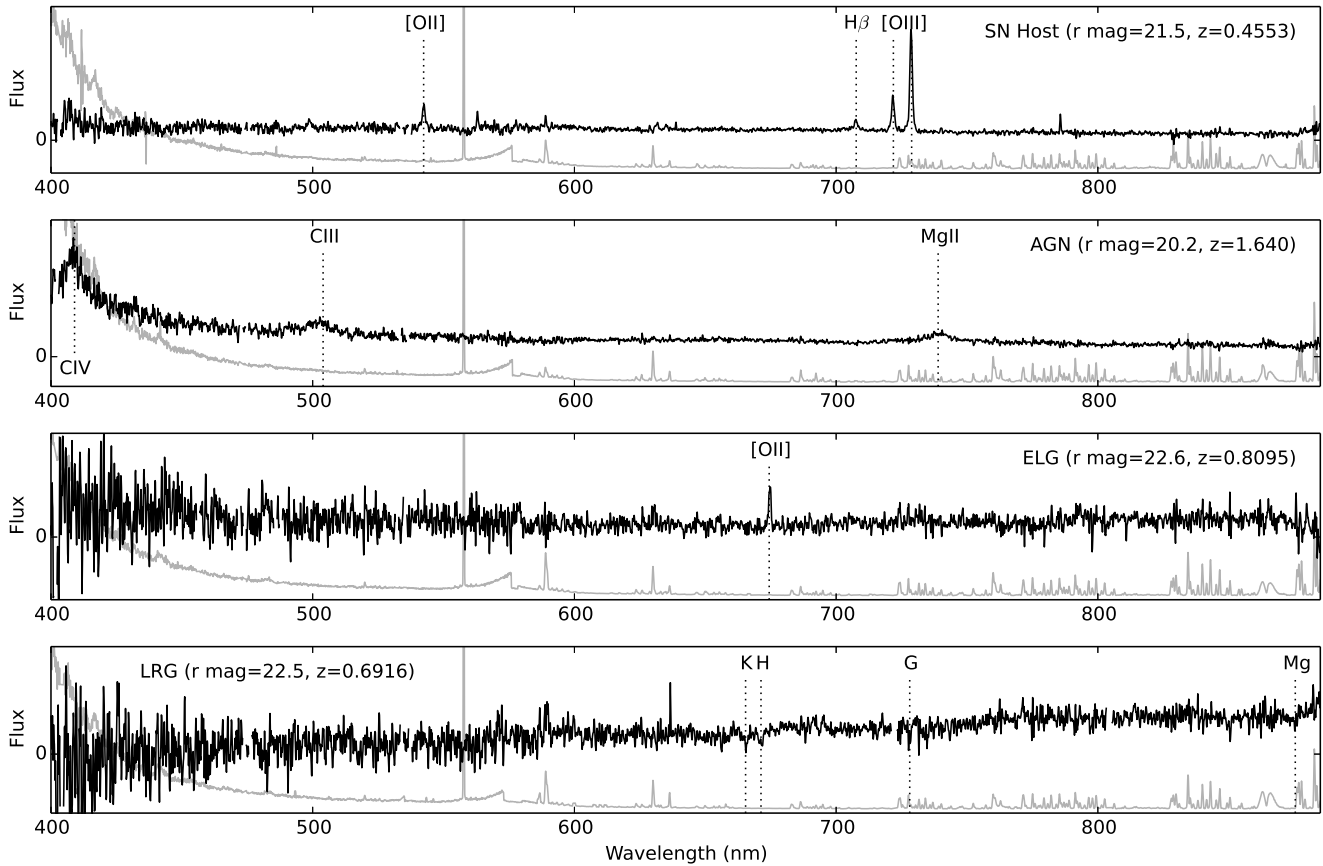}
\caption{Sample spectra for targets of typical magnitude and redshift, with flux on a linear scale.
Spectra are truncated at the blue end and smoothed by a gaussian filter of sigma 1 pixel. 
Key features used to measure the redshifts are marked.
The spectra are not flux calibrated and the relative variance is plotted in gray. The discontinuity at 570~nm in the variance marks the dichotic split. }
\label{fig:sample_specs}
\end{minipage}
\end{figure*}

The number of each type of target we observed by end of Y1 is
given in Table~\ref{tab:target_type}. 
Examples of spectra for the main types of targets are shown in Fig.~\ref{fig:sample_specs}.

\begin{table*}
\centering
\begin{minipage}{\textwidth}
\caption{OzDES Target Types and Priorities.  Highest priorities are
  allocated first, up to their quota. 
  Objects are removed from the target
  pool based on the deselection criteria.  In most cases this is when
  the target has a successful redshift measurement, but for transients
  (which include supernovae) it is after they have faded, and some are
  never deselected as they require constant monitoring (AGN and F
  stars). The final column shows both the number with successful redshifts, and the redshift success rate as a percentage of the number observed.  
  }

\label{tab:target_type}
\begin{tabular*}{\textwidth}{@{\extracolsep{\fill}}lcccccc}
\hline
Type	    & Priority& Quota      & Deselection & Number     & Average Exposure &  Number \\
            & (1-9)   &            & criterion   & Observed   & (minutes)        & With Redshift  \\
\hline

Transient         &  8&  unlimited & faded      &         327 &        192&        238 (73\%)\\
AGN reverberation &  7&        150 & redshift or never \textsuperscript{*}   &        2103 &        283&       1772 (84\%)\\
SN host           &  6&        200 & redshift   &         986 &        194&        528 (54\%)\\
WhiteDwarf        &  6&          3 & classification   &         17  &        139&         ...      \\
StrongLens        &  6&          3 & redshift   &         15  &        181&          3 (20\%)\\
ClusterGalaxy I   &  6&         10 & redshift   &        439  &        165&        232 (53\%)\\
RadioGalaxy I     &  6&         25 & redshift   &        350  &        259&        161 (46\%)\\
FStar             &  5&         15 & never\textsuperscript{$\dagger$}      &         48  &         88&         ...      \\
Sky Fibres        &  5&         25 & never      &         ..  &	...	&	...	\\
ClusterGalaxy II  &  4&         50 & redshift   &        133  &        249&         89 (67\%)\\
LRG               &  4&         50 & redshift   &       1208  &        157&        728 (60\%)\\
ELG               &  4&         50 & redshift   &       2382  &        156&       1326 (56\%)\\
Photo-$z$         &  2&  unlimited& redshift    &       2192  &        123&       1621 (74\%)\\

\hline
\\
\multicolumn{7}{l}{\textsuperscript{*}\footnotesize{Never deselected if a target is picked to be monitored.}} \\
\multicolumn{7}{l}{\textsuperscript{$\dagger$}\footnotesize{Never deselected if confirmed as an appropriate calibration source.}}\\

\end{tabular*}

\end{minipage}
\end{table*}

\begin{enumerate}

\item \textit{Transient}: transient of any type, including
  supernova. Currently, an approximate $r$-band magnitude limit of
  22.5, corresponding to the peak r-band magnitude of a Type Ia
  supernovae at $z\sim0.5$, is imposed on the brightness of
  the transient. This magnitude is the limit at which we can
  identify a hostless Type Ia supernova with AAOmega. Because of the
  time-sensitive nature of transients, these objects are of the
  highest priority. Regardless of whether a redshift or a classification is obtained, a transient remains on the target list until it becomes fainter than the magnitude limit.
  There are only 5 to 10 active transients at any
  one time in each DES field, so all of these will be assigned a fibre
  unless a collision with another transient occurs.

\item \textit{AGN Reverberation}: AGN candidate or previously-identified AGN that we began to monitor in Y1.
The AGN candidates were selected based on photometry from DES, VHS \citep{2013Msngr.154...35M}, 
and WISE \citep{2010AJ....140.1868W} with a variety of methods \citep{2015MNRAS.446.2523B}. 
The selection methods were designed to emphasize completeness over efficiency, 
particularly for the brightest candidates, in order to maximize the number of bright AGN. 
Besides brightness, we use redshift, emission-line equivalent width, and luminosity to identify targets for monitoring. 
For each run in Y1, approximately 150 fibers per field were devoted to AGN targets (Table 2).
As Y1 progressed, a growing fraction of these fibers were devoted to the monitoring program.

\item \textit{SN Host}: host galaxy of a supernova or other
  transient detected since the DES SV season. A host galaxy is
  defined as the closest galaxy in units of its effective light radius
  from the SN \citep[cf.][]{2014arXiv1401.3317S,
    2006ApJ...648..868S}. An allocated fibre is placed at the core of the galaxy to maximize the flux input (but see $\S$~\ref{sec:hostless} for a complementary strategy employed in Y2).
   Around 20 new hosts are
  identified for each DES field for each OzDES observing run. The
  total number of targets accumulates as fainter objects remain in the
  list until it becomes clear that we are unlikely to get a
  redshift. During the first year, no target was dropped. We are
  currently using these data to assess when targets should be dropped
  in favour of new targets.

\item \textit{White Dwarf}: white dwarf candidate. The list is updated between observing runs based on analysis of the observed spectra,
when successfully classified targets are dropped.

\item \textit{Strong Lens}: strong lens system identified in the DES imaging. Up to five lens systems per field were targeted in Y1.

\item \textit{Cluster Galaxy}: cluster galaxy selected via the
  redMaPPer cluster finder on the SVA1 Gold catalog \citep[a galaxy catalog from co-added SV imaging;][Rykoff et
    al., in prep]{2014ApJ...785..104R}.  High probability cluster galaxies are
  selected to be luminous galaxies in moderately rich clusters that
  have a luminosity consistent with the cluster richness, as well as
  occupying regions with a high local density.  An additional
  constraint of $m_r<22.5$ mag is put on the galaxies for AAOmega
  targeting. Central cluster galaxies with $0.6 < z_\mathrm{photo} <0.9$ are given a
  higher priority (ClusterGalaxy I in Table~\ref{tab:target_type}),
  with the median target redshift of $0.80$.  Another lower
  priority list of targets (ClusterGalaxy II) is made for centrals and
  bright satellites of clusters at $z<0.6$.

\item \textit{Radio Galaxy}: galaxy selected from the ATLAS survey.
  Although these are labelled ``Radio Galaxy'' for convenience, about half the objects are star forming galaxies or radio-quiet AGN.
  As for cluster galaxies, we allow two priority settings, for RadioGalaxy I and
  RadioGalaxy II. During Y1, category II was rarely used (only two objects), so it
  is not listed in Table~\ref{tab:target_type}. 
  
\item \textit{F star}: F star candidate in Y1 (from Nov. 2013), to
  obtain spectroscopic classification.  The goal is to have
  approximately 10 to 15 high signal to noise F stars per field to be
  used as flux calibrators.  Candidates were selected in the Northern
  fields using photometry from SDSS that included u-band.

\item \textit{Sky}: sky regions with no detectable sources in DES
  imaging. Approximately 25 fibres are allocated per field.

\item \textit{LRG}: luminous red galaxy used to calibrate DES
  photometric redshifts.  The high-$z$ (median$\sim$0.7) population is
  selected based on DES+VHS photometric data
  \citep{2015MNRAS.446.2523B}.
 
\item \textit{ELG}: emission line galaxy used to calibrate DES
  photometric redshifts. Color selections similar to \citet{2013MNRAS.428.1498C} are used.
For bright targets ($19 < m_i < 21.3$), $-0.2 < (m_g - m_r) < 1.1$ and $-0.8 < (m_r - m_i) < 1.4$;
for faint targets ($21.3 < m_i < 22$),  $-0.4 < (m_g - m_r) < 0.4$, $-0.2 < (m_r - m_i) < 1.2$ and $m_g - m_r < m_r - m_i$.

\item \textit{photo-$z$}: galaxy target, selected using the DES i-band 
magnitude cut $17 \leq m_i < 21$, used to calibrate DES photometric redshifts.

\end{enumerate}

\subsection{Data Reduction}\label{sec:reduction}

We process the data from AAOmega soon after they have been taken so
that we are able to quickly determine redshifts, usually within 48
hours of them being observed and often before the next night's
observations.  This gives us the chance to deselect targets if a
secure redshift (see below) has been obtained and therefore free up
fibres to observe other targets later in the run.

All data from AAOmega are processed with \texttt{2dfdr} \citep{Croom2004}. The procedure is
broken into a number of steps, each of which is discussed below.

\begin{itemize}

\item {\it Bias Subtraction and Bad Pixel Masking}. For data taken
  with the blue CCD, this consists of using the overscan region to
  subtract the bias, subtracting a master bias to remove features that
  cannot be removed by using a fit to the overscan region, and
  subtracting a scaled version of the master dark. For the red CCD,
  this simply consists of using the overscan region to remove the
  bias. Cosmic rays are detected with an edge detection filter. The affected pixels
  are masked as bad throughout the subsequent analysis and are not
  used. Some cosmic rays sneak through this step, but are captured
  later when multiple exposures are combined.

\item {\it Tram line mapping and wavelength calibration}. The fibre
  flat is used to measure the location of the fibre traces (the tram
  line map), while the arc is used to wavelength calibrate the path
  along each fibre. In future versions of \texttt{2dfdr}, the fibre flat will
  also be used to model the fibre profile.

\item {\it Extraction}. A 2d spline model of the background scattered
  light is then subtracted, and the flux from individual fibres is
  extracted. The flux perpendicular\footnote{In practice we weight
    along the direction that is parallel to the detector columns,
    which is almost perpendicular to the fibre traces.} to the fibre
  trace is weighted with a Gaussian that has a FWHM that matches the
  FWHM of the fibre trace.  In future versions of \texttt{2dfdr}, both of these
  steps will be replaced with a single step that optimally extracts the
  flux and determines the background \citep{2010PASA...27...91S}.

\item {\it Wavelength calibration}. The extracted spectra are
  calibrated in units of constant wavelength using the arcs.

\item {\it Sky subtraction}. The relative throughput of the fibres is
  normalised, and the sky is removed using the extracted spectrum of
  the sky fibres. Usually, there are residuals that remain after this
  step. These are removed using a principal component analysis of the
  residuals \citep{2010MNRAS.408.2495S}.

\item {\it Combining and Splicing}. If more than one exposure was
  obtained, which is usually the case, the reduced spectra are
  co-added. Remaining cosmic rays are found as outliers and removed at
  this stage.  The red and blue halves of the spectra are then
  spliced. At this stage, we do not weight the data before we combine it. In future
  versions of the OzDES pipeline, we will weight the spectra and both
  the splicing and combining will be merged into a single step.

\end{itemize}

\subsection{Redshifting}\label{sec:redshifting}

All spectra are visually inspected by OzDES team members\footnote{A
  person who determines redshifts using \texttt{runz} is colloquially referred
  to as a redshifter.} using the interactive redshifting software
\texttt{runz}, originally developed by Will Sutherland for the
2dFGRS. \texttt{runz} first attempts an automatic redshift
determination by both cross-correlating to a range of galaxy and
stellar templates and searching for emission line matches. It then
displays the spectrum and marks the locations of the common emission
and absorption features at the best redshift estimate. A redshifter 
visually inspects the fit and determines whether the redshift is
correct. A number of interactive tools are provided, e.g. to switch to
a different template, to mark a specific emission line, to add a
comment, or to input the redshift directly.

Uncertainties on the redshifts are calculated by \texttt{runz} based
on the width of the cross-correlation peak or the fit to emission
lines.  However, we are able to derive more representative
uncertainties for various classes of objects (e.g.~galaxies and AGN, see~\ref{sec:precision})
using objects that are observed multiple times. 

Each redshift is assigned a quality flag. For most objects, we use a
number between 1 and 4, with a larger number meaning a more
secure redshift estimate. We reserve quality flag 6 for stars.

\begin{itemize}
\item Quality 4 is given when there are multiple strong spectral line
matches.  With the exception of AGN and transients, these are removed
from the target pool and are not re-observed.

\item Quality 3 is given for multiple weak spectral line matches or single
strong spectral line match (e.g. a bright emission line that is consistent with high-redshift [OII]).  These
can be used for science, but for some target types (e.g. SN hosts), may also be
re-observed until the quality is deemed worthy of a 4.

\item Quality 2 is given to targets where there are one or two very weak
features (e.g. a single weak emission line that may be [OII]).  The redshift is speculative and
not reliable enough for science.

\item Quality 1 is given to objects where no features can be identified.
\end{itemize}

Targets with quality 1 and quality 2 redshifts are re-observed
until the deselection criteria are met.

In practice, the assignment of quality flags by a redshifter is
subjective to experience and many other factors. We require
independent assessments from two members of the team for each
object. The results from the two redshifters are then compared by a third
``expert'', who chooses the appropriate redshifts and quality flags and provides feedback to the individual
redshifters.
This helps to train redshifters and to homogenise the
redshifting process. The number of cases where there is true
disagreement is small, most disagreements arise from a difference in
quality rating.

At the beginning of Y1, we set the following requirements
for the reliability of the redshift:

\begin{itemize}
\item[] 6: more than 99\% correct (reserved for stars)
\item[] 4: more than 99\% correct (reserved for galaxies)
\item[] 3: more than 95\% correct (for any object)
\end{itemize}

Redshifts with quality of 3 and above are considered 
trustworthy for science analysis, so it is important to know the
actual rate of redshift blunders. 
By comparing objects that were observed multiple times and redshifted
independently, we find that we are achieving the required level of
reliability for quality flags 4 and 6, but are tracking below 95\% for
quality 3. These results are presented in $\S$~\ref{sec:reliability}.

\subsection{Classification of Active Transients}\label{sec:transient_classification}

Timely identification is critical for transient studies. Discovery of
a supernova at an early phase may trigger observations at other
wavelengths or spectroscopic time series throughout its evolution. The
earlier the follow-up campaign starts, the more information it will
gather to understand the explosion physics. In the exciting case of an
unknown type of transient, early observations may be vital for the
interpretation of its true nature and the only chance to observe it if it is short-lived.

Spectral classification of a SN is usually done by comparison to
templates. The best match type and age are found through either
cross-correlation \citep[e.g. the Supernova Identification,
  \texttt{SNID},][]{2007ApJ...666.1024B} or chi-square minimization
\citep[e.g. \texttt{Superfit},][]{2005ApJ...634.1190H}.

During the first season of OzDES, we observed 320 transients and
determined the redshifts (mostly from features of the host galaxies) of
about 73\% of them. Of these, 12 were positively identified as supernovae
\citep{2013ATel.5568....1C,2013ATel.5642....1Y,2014ATel.5757....1Y}. This
is a small fraction of all transients. There are a number or reasons
for this.

First, one of the aims of OzDES is to explore the range of transient
phenomena that exists in the DES deep fields, so a deliberate
decision was made to target objects that were clearly not SN. This
includes objects that turned out to be AGN or variable stars.

Second, the amount of host galaxy light relative to SN light that
enters the $2\arcsec$ 2dF fibre is larger than is normally the case for long slit
observations, which typically use slits that are $1\arcsec$
wide. This, coupled with the typical seeing at the AAT ($2\arcsec$), 
makes the SN less clearly visible in spectra from 2dF.

Third, SN typing relies on identifying broad spectral features. At
the beginning of Y1, the pipeline that was used to process the data
imprinted features to the spectra (e.g. a discontinuity between the
red and blue halves) that obfuscated the SN signal. Only the brightest
SN in Y1 could be identified with confidence.  

\subsection{Ongoing improvements}\label{sec:y2_upgrades}

While the Y1 spectra are suitable for determining
redshifts, artefacts that come from the processing result in data that
are less suitable for other analyses, such as the spectral typing of
transients.  The most common artefact is a flux discontinuity in the
overlap region between the red and blue halves of the spectra.

Since the first year of the OzDES campaign, considerable work has
gone into mitigating these artefacts by improving the algorithms in
\texttt{2dfdr}.  These improvements include better tramline tracking,
implementation of optimal extraction (which leads to more accurate
treatment of the background scattered light), better flat fielding, 
more effective cosmic ray removal and improved flux
calibration. During Y1, we have used the sensitivity functions provided with \texttt{2dfdr} to do the flux calibration. In future versions of the OzDES pipeline, we will
use the F stars that are observed contemporaneously with the other targets to do the flux calibration. Absolute calibration will be done using the
broad band DES photometry.

Not all of the above algorithms are implemented at once, but better data reduction has already helped us to classify 50\% more SNe in Y2 than in Y1. 
Reprocessing of the entire OzDES data set from Y1 and Y2 is planed, after which redshifts and spectra will be publicly released (Childress et al. in prep).

\section{Results}\label{sec:results}

\begin{figure*}
\begin{minipage}{\textwidth}
\includegraphics[ width=170mm]{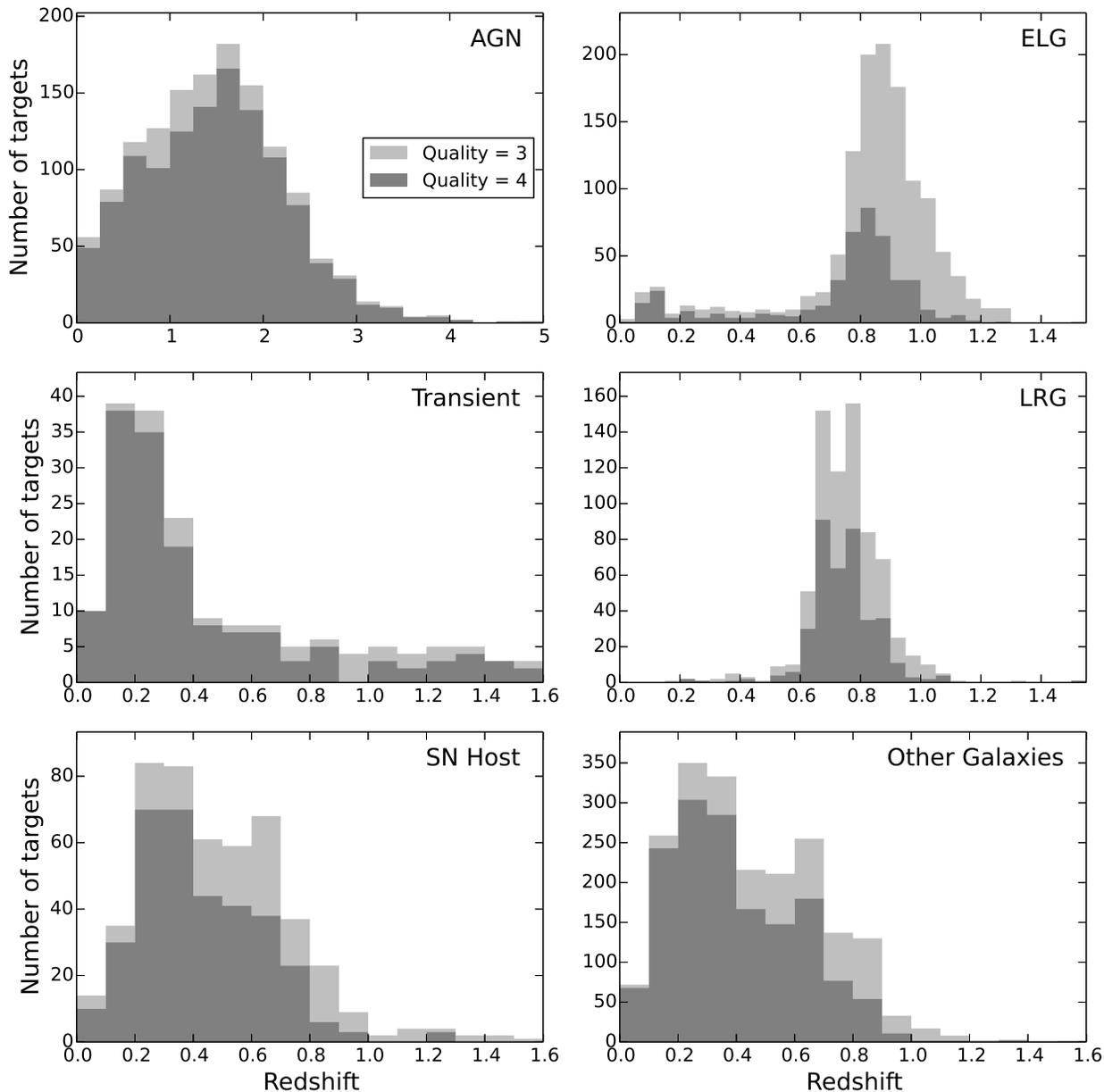}
\caption{Redshift distributions for selected target types. Objects that have redshifts consistent with Galactic origins are excluded. 
Dark shaded histograms represent objects with redshift quality flag of 4 (most reliable; see $\S$~\ref{sec:redshifting}) and the light shaded histograms represent objects with redshift quality flag of 3 (reliable).
Other galaxies include radio galaxies, cluster galaxies and galaxies that are specifically chosen to calibrate photometric redshifts.
Redshift bins are set to be smaller for ELGs and LRGs to provide higher resolutions around the peak.}
\label{fig:hist_rz}
\end{minipage}
\end{figure*}

\begin{figure*}
\begin{minipage}{\textwidth}
\includegraphics[ width=170mm]{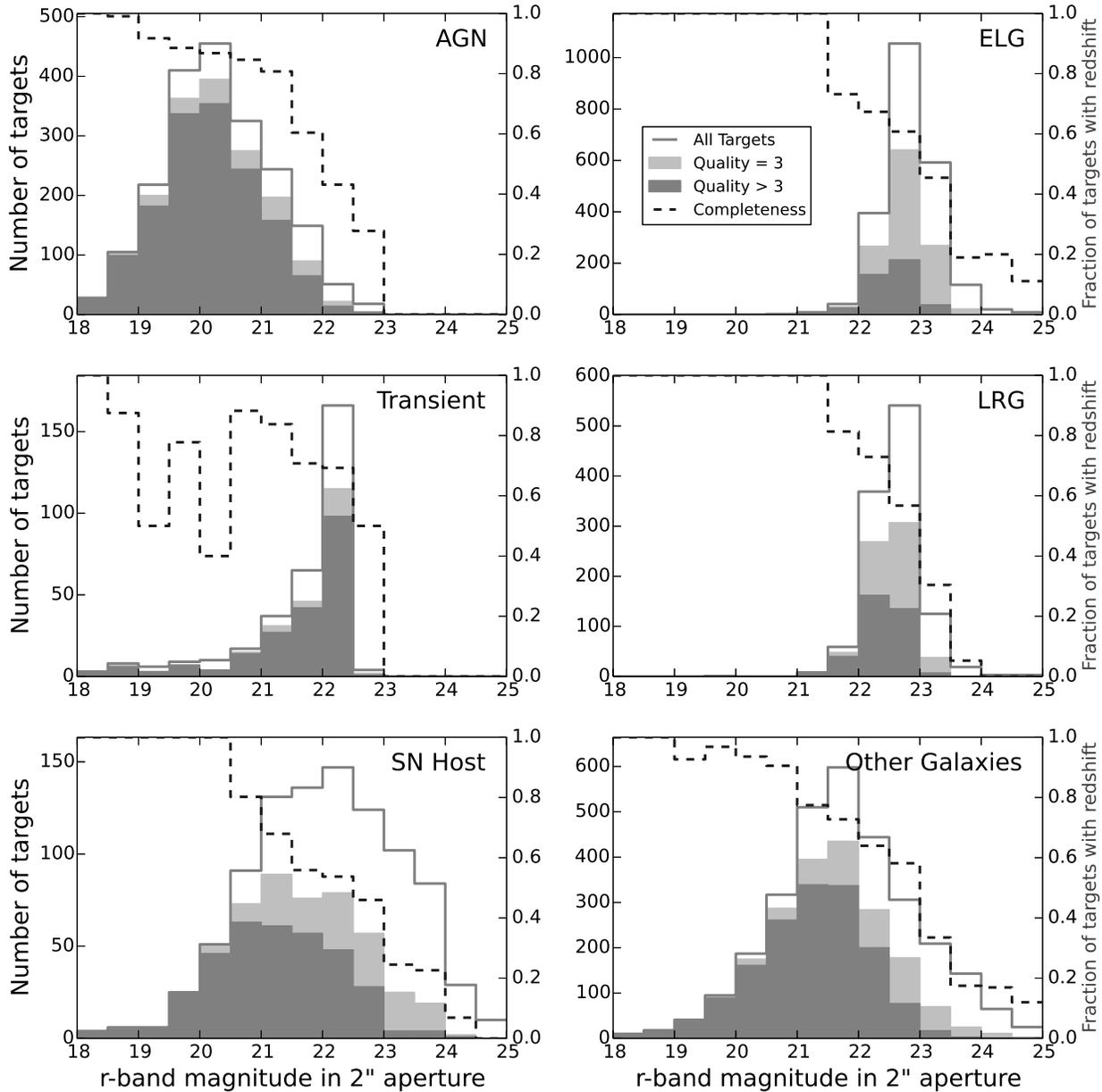}
\caption{Redshift completeness as functions of r-band magnitude measured in a 
$2\arcsec$ diameter aperture, for selected groups of extragalactic
  objects. Unfilled histograms are for all targets. 
  Dark shaded histograms represent objects with redshift quality of 4 and above (most reliable; see $\S$~\ref{sec:redshifting}). 
  Light shaded histograms represent objects with redshift quality of 3 (reliable). 
  Completeness (as dashed curves) is defined as the fraction of objects for
  which redshifts are measured with quality flag 3 and above.
  The magnitude range is fixed for all panels for easy comparison. 
  A few galaxies in the bottom two panels have measured magnitudes fainter than 25. These bins have low completeness and are excluded to emphasize the more typical magnitude range.}
\label{fig:hist_comp}
\end{minipage}
\end{figure*}

During Y1 and SV, 10482 unique targets were observed and 6727
redshifts (with quality flag 3 and above) were obtained. Figure
\ref{fig:hist_rz} shows the redshift distributions for selected
targets. The fraction of objects with measured redshifts and their
number are listed in Table~\ref{tab:target_type} for each type of
target. For selected groups of extragalactic objects, we also show the
completeness of redshifts as a function of integrated magnitude within
the fibre diameter (Figure~\ref{fig:hist_comp}). For non-transient
objects, photometry measurements are taken from the DES SV Gold
catalog (Rykoff et al., in prep). For transients, the magnitude does not include the
flux of the underlying host and represents luminosity measured in the last epoch when the target was selected.

\subsection{Efficiency and Completeness}\label{sec:efficiency}

As expected, the probability of a successful redshift measurement
drops for fainter targets. This trend is weak for the transients 
because the redshifts of most transients are
inferred from spectral features of their hosts. 
Strength of such features depends on the host luminosity, redshift and location of the transient relative to its host.

The redshift efficiency is the likelihood of measuring a redshift
above a certain quality level in a given observation time. Difference
in completeness during comparable integration times reflects difference
in efficiency for different types of targets. In general, emission
lines allow measurements of redshift for sources with faint continuum,
while cross-correlation for a galaxy without identifiable emission
features relies on well measured spectral shape, combined with
absorption features that are usually relatively weak, and therefore
needs brighter continuum. Hence, for ELGs, the efficiency is less
sensitive to the integrated broad-band magnitude than for the LRGs.

In practice, the ELGs are selected to be at $z\sim1$, so only the [OII] $\lambda\lambda$ 3726,3728 doublet is within our wavelength
range. This double line appears blended at the resolution of our
spectra, and although it can often be recognized as a wide or
flat-topped line, it is a single feature that is prone to
misidentification.  It is hard for redshifters to say conclusively
that it is [OII] and not another line or an artefact. A high fraction
of the ELGs are therefore rated at a redshift quality of 3. 

The nominal target type is known to the redshifter. For ELGs, LRGs and AGN,
this may cause the redshifter to assign a higher redshift quality flag
than would be the case if the target type was not available to the
redshifter.

SN hosts show
lower completeness across a wide magnitude range than either ELGs or LRGs 
(see Figure~\ref{fig:hist_comp}). This sample is selected based on criteria
independent of whether the galaxy is of early or late type. The
redshift range may be constrained by the selection criteria, but
knowing the target type does not directly help in determining a
redshift.

At a given magnitude, the redshift completeness grows with integration
time. Many of the faint targets are expected to be observed in
multiple seasons and accumulate significantly more signal than
obtained so far.
The data from Y1 do not have sufficient range of exposure time to determine this trend with much certainty.
In O'Neill et al. (in prep), we use data gathered by the end of Y2 to model how the number of redshifts acquired
increase as a function of number of visits to the field, for various types of targets with different magnitude and redshift distributions.
We then estimate the redshift completeness of the survey
and the number of objects we are likely to observe by extrapolating to a total of 25 visits. 
We are on track to obtain
redshifts for 80\% of all SN host galaxies that are targeted.

\subsection{Redshift Precision}\label{sec:precision}

Estimates of the redshift precision based on emission line fitting or
template cross-correlation may suffer different biases. For almost all
of our target groups, a combination of techniques has been used. It
is recognized that for most of our science cases, we exceed the
minimum requirement of precision. Therefore, we do not attempt to
assign accurate uncertainties for individual redshift measurements,
but instead provide an overall statistical error estimate for classes
of targets, based on the dispersion in measured redshifts for
individual objects with multiple independent measurements, either from
different observing runs or from appearing in the overlap regions of
the DES fields. Objects with inconsistent redshift measurements are excluded (see next section).
For each object, we calculate the pair-wise
differences of these redshifts, divided by $1+z_{\rm median}$. The distributions
are examined for several populations of extra-galactic objects. We
quote the 68 percent ($\sigma_{68}$) and 95 percent ($\sigma_{95}$) inclusion regions in Table
\ref{tab:error}, since the distributions have extended tails. Typical uncertainties ($\sigma_{68}$) for AGN and galaxies are 0.0015 and 0.0004 respectively.

We also compare our results against the redshifts from other surveys covering our targeted sky area. 
Only redshifts with the most confident quality flag defined by the corresponding survey are considered.
The results are listed in Table \ref{tab:error} as
``external''. No significant systematic offset is found between OzDES and surveys such as GAMA \citep{2011MNRAS.413..971D} or SDSS \citep{2014ApJS..211...17A}. 
Distributions of the differences between our redshifts and those
from other surveys are consistent with the internal
comparisons.

A large number of AGN were targeted repeatedly from run to run by
design. This allows us to split the sample and quote the uncertainties
in two redshift bins. The redshifts of AGN are often measured by
cross-correlating with template spectra. Due to intrinsic variation of
the emission profiles from AGN to AGN, the precision of a measurement
depends on the quality of the template, the wavelength region
included in the cross-correlation and/or the line chosen by the redshifter
to centre on. At redshift one and above, high ionization emission
lines with large profile variation, such as CIII] or CIV are often
  used, resulting in larger uncertainties in the redshifts.

The number of repeated observations is small for all other galaxy
targets as they are usually removed when secure redshifts are
obtained. To investigate possible type dependence, we examine
separately the dispersions for ELG, LRG and SN host targets. The
smaller dispersion for the ELGs is consistent with the expectation
that better constraints can be obtained from narrow emission lines.

\subsection{Redshift Reliability}\label{sec:reliability}

For a quantitative evaluation of the reliability of our redshifts, we
compare multiple redshifts obtained for the same objects from 
independent observations, either from
different observing runs or from appearing in the overlap regions of
the DES fields. For each redshift, we calculate its offset from a redshift
that is deemed to be correct. For simplicity, we call this value the
base redshift. We elaborate on how the base value is determined below.
A redshift is considered wrong if it differs from the base redshift by more than 
$0.02 (1+z)$ for a AGN or $0.005 (1+z)$ for a galaxy, which are roughly 
ten times the standard deviations measured from the previous section. 

The base redshift can be obtained from either internal or external
sources. Internally, if a redshift with a higher quality flag exists,
we consider that redshift as the base redshift. Otherwise, the median
value is used if more than two redshifts are available or a random
selection is made assuming at least one measurement is correct. 
The external redshifts are chosen from other surveys to have the highest quality flag defined. 
When a conflict between an OzDES redshift and an external redshift is found, 
the OzDES spectra are re-examined. In three cases (for quality 3 AGN), 
the data quality is not good enough to confirm either redshift. These are excluded in the calculations.

As shown in Table \ref{tab:error}, overall close to 100\% of redshifts
are correct with quality flag 4 but only about 90\% of redshifts are
correct with quality flag 3. 
To better understand the sources of error, we examine different galaxy
types separately. 
The relatively high error rate for SN hosts (more than 15\% for quality 3)
possibly arises because of the diversity of objects that host
supernovae. The ELG sample appears more homogeneous. 

As noted earlier, the error rate for quality 3 objects is higher than 
our goal of 5\%. After Y1, we implemented a
number of changes (better training of the human redshifters and more
scrutiny of quality 3 objects by the third person) that have
resulted in fewer errors.

With increasing experience and
better data processing, further
improvements on the reliability and the quality of the
redshifts are expected in the coming seasons and will be closely
monitored.

\begin{table*}
\centering
\begin{minipage}{\textwidth}
\caption{Redshift uncertainties and error rates. Uncertainties are calculated using weighted pair-wise redshift differences, $\Delta z/(1+z)$, for objects observed in multiple overlapping fields or multiple observing runs ($\S$~\ref{sec:precision}). A redshift is considered wrong if it differs from a chosen base redshift ($\S$~\ref{sec:reliability}) by more than 
$0.02 (1+z)$ for a AGN or $0.005 (1+z)$ for a galaxy.}
\label{tab:error}
\begin{tabular*}{\textwidth}{@{\extracolsep{\fill}}cccccc}
\hline
Type	 &   Number of redshift pairs & $\sigma_{68}$\textsuperscript{*} & $\sigma_{95}$\textsuperscript{*}  & Error rate (Q = 4) & Error rate (Q = 3) \\

\hline
AGN                & - & - & -                     & 1/1647 (0.1\%) &  22/388 (5.7\%) \\
AGN ($z\leq1$) & 521   & 0.0004 & 0.0011 & - & - \\
AGN ($z>1$) & 1568 & 0.0015   & 0.0038    & - & -\\

AGN (external) & 424 & 0.0016 & 0.0048 & 0/387 (0.0\%) \textsuperscript{$\ddagger$} 
& 6/46(13.0\%)\textsuperscript{$\S$} \\

Galaxy\textsuperscript{$\dagger$}
                 & 99   & 0.0004 & 0.0013 & 0/74 (0.0\%)   &  10/95 (10.5\%) \\
ELG         & 25   & 0.0002 & 0.0006 &  0/8 (0.0\%)   &  2/32 (6.3\%)  \\
LRG         & 21   & 0.0005 & 0.0013 &  0/16 (0.0 \%)  &  3/20 (15.0\%)  \\
SN host   & 36   & 0.0003 & 0.0013 &  0/38 (0.0\%)   & 4/25 (16.0\%) \\

Galaxy (external) 
                 & 182 & 0.0004 & 0.0010 & 0/159 (0.0\%)\textsuperscript{$\parallel$} & 0/24 (0.0\%) \\
\hline
\\

\multicolumn{6}{l}{\textsuperscript{*}\footnotesize{68 percent ($\sigma_{68}$) and 95 percent ($\sigma_{95}$) inclusion regions.}} \\
\multicolumn{6}{l}{\textsuperscript{$\dagger$}\footnotesize{All types of galaxy targets that are not AGN.}} \\
\multicolumn{6}{l}{\textsuperscript{$\ddagger$}\footnotesize{In three cases of discrepancy, spectra are inspected and the OzDES measurements are confirmed to be correct.}}\\
\multicolumn{6}{l}{\textsuperscript{$\S$}\footnotesize{All discrepant redshifts are checked by hand. Objects are excluded if the available data is not good enough to confirm either}} \\
\multicolumn{6}{l}{\footnotesize{redshift measurement.}} \\
\multicolumn{6}{l}{\textsuperscript{$\parallel$}\footnotesize{In one case of discrepancy, spectra are inspected and the OzDES measurement is confirmed to be correct.}}  \\

\end{tabular*}
\end{minipage}
\end{table*}

\section{Discussion and highlights}\label{sec:discussion}

In this section, we present updates on some of our key science goals in light of the results from Y1. When necessary, complementary data taken in Y2 are included to demonstrate the potential of our observing strategy. Some of the discussions involve changes in Y2 that are inspired by analysis of the Y1 data.

\subsection{SN Ia Cosmology}\label{sec:cos_disc}

The DES SN survey strategy is optimized to detect a large number of SNe Ia with a redshift distribution that extends to redshift 1.2 and peaks around redshift 0.6 \citep[cf. Figure~10 of][]{2012ApJ...753..152B}. Since host galaxy redshifts from OzDES will be the main source of redshifts for Hubble Diagram analyses, the efficiency of obtaining redshifts by OzDES will affect the actual redshift distribution in addition to the discovery efficiency of DES.

\begin{figure}
\includegraphics[width=84mm]{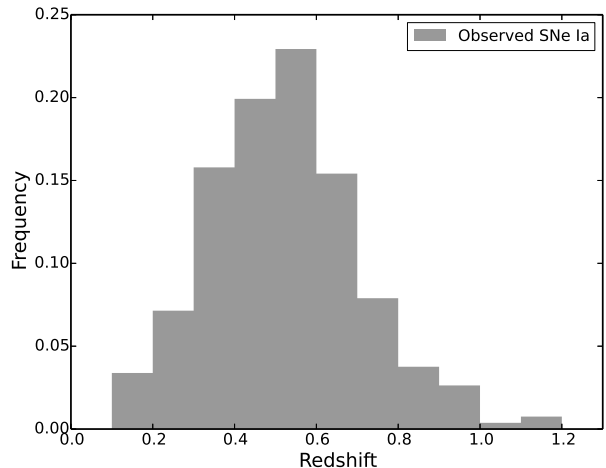}
\caption{Normalized redshift distribution of selected SNe Ia candidates from Y1 that have host spectroscopic redshifts. The selection criteria are tuned to achieve $>$ 96\% purity (see text in $\S$~\ref{sec:cos_disc}). The median redshift of this sample is 0.52.}
\label{fig:comparison_Ia}
\end{figure}

In practice, SNe Ia candidates are first selected based on DES light-curves and photo-$z$ of putative host galaxies. The host galaxies of these candidates are then targeted by OzDES. Any redshifts obtained are subsequently used to refine the selection. Only those SNe Ia with a secure host spectroscopic redshift are considered further for cosmology studies. The combined efficiency of this process can be evaluated by comparing the selected SNe Ia sample to a simulated population. 
From the DES Y1 data, we select SNe that satisfy light curve quality cuts defined in Table~6 of \citet{2012ApJ...753..152B}, with the exception that only two filters are required to have maximum signal-to-noise above 5 (i.e. 2 filters with SNRMAX $>$ 5). The looser cuts are compensated for by additional selection criteria on peak color and \texttt{SALT2} \citep{2007A&A...466...11G} model fit parameters ($x1$ and color) to achieve $>$ 96\% purity \citep[tuned for simulation, cf.][]{2013APh....42...52G}. 
We show in Figure~\ref{fig:comparison_Ia} the redshift distribution of the selected SNe Ia for which host spectroscopic redshifts are available by end of Y2. 
The median redshift of this observed sample is 0.52, compared to a median redshift of 0.63 for SNe Ia selected with the same photometric criteria from a \texttt{SNANA} \citep{2009PASP..121.1028K} simulation with realistic survey characteristics (e.g. observing cadence, seeing conditions and photometry zero points) from Y1 as input. 
The lower median redshift for the observed sample is mainly due to the difficulty of measuring redshifts in more distant and fainter galaxies. 
 In the two deep fields (C3 and X3), half of the SNe Ia host galaxies are fainter than 24th magnitude in r-band and the overall redshift completeness is only about half of that for the shallow fields.

With longer exposure times, the redshift completeness will improve at all redshifts. Only for the most distant SNe, spectroscopic redshifts may remain scarce for host galaxies at typical brightness. This bias against faint host galaxies needs to be considered as it has been shown that stretch and color-corrected SN Ia peak magnitude depend on the host galaxy stellar mass \citep[e.g.][]{2010ApJ...715..743K, 2010MNRAS.406..782S, 2010ApJ...722..566L, 2013ApJ...770..108C}.

\subsection{Faint SN hosts and Spectra Stacking}
\begin{figure*}
\begin{minipage}{\textwidth}
\includegraphics[width=180mm]{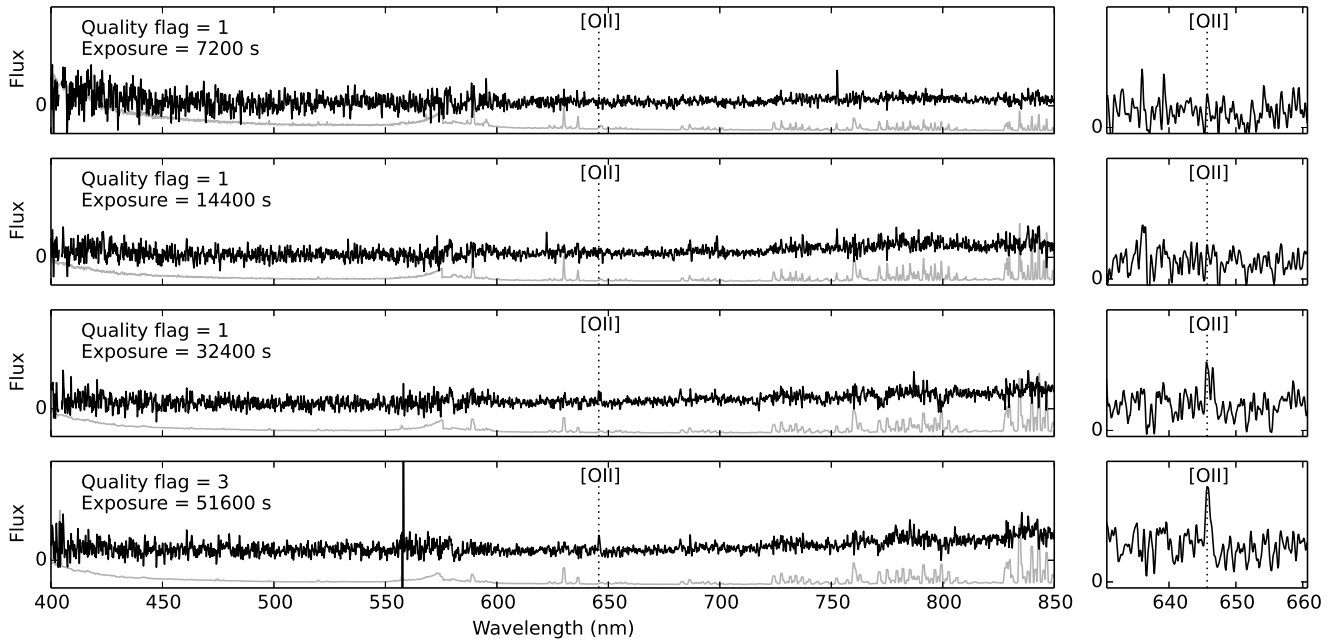}
\caption{Selected co-added spectra of a SN host target of r-band magnitude 23.7 and redshift 0.732. The significance of the candidate [OII] feature appears to have increased with exposure time. A quality flag of 3 is assigned so the target will remain in the queue until more data confirms the redshift.}
\label{fig:faint_spectra}
\end{minipage}
\end{figure*}

As discussed above, high redshift efficiency across a wide brightness range is critical to maximize the statistics and minimize the bias of a sample of SNe with host galaxy spectroscopic redshifts. Such a goal is achieved by OzDES's unique strategy of repeat targeting, dynamic fibre allocation, and stacking. Multiple observations of the same DES field allows dynamic control of effective exposure times for targets of different brightness. The brighter galaxies are deselected between the observing runs when redshifts are measured; while fainter galaxies remain in the queue. Stacking across many observing runs allows OzDES to go deeper than otherwise possible with the 3.9m AAT. Figure~\ref{fig:faint_spectra} shows an example of stacked spectra, for a SN host target of r-band magnitude 23.7 and redshift 0.732. The growing significance of the emission line, roughly consistent with square root of exposure time, supports the credibility of the feature.

\subsection{Hostless SNe and Super-Luminous SNe}\label{sec:hostless}

We showed in $\S$~\ref{sec:efficiency} that redshift efficiency
has a different dependence on target brightness for different types of
galaxies. The likelihood of measuring an emission line redshift is
less sensitive to apparent broad-band luminosity. A significant
fraction of SNe (including most SNe Ia and most core-collapse SNe) are
expected to occur in star-forming regions. For some, strong emission
lines may show up in dispersed AAT spectra even though the continuum
is too faint to be detected by DECam imaging in the first few seasons. A new strategy of
targeting hostless SN, by placing a fibre at the position of a SN after
the transient has faded, is implemented after Y1 and has already yielded
redshifts that would otherwise be elusive.

This strategy is particularly interesting for validating Super-Luminous SN (SLSN) candidates. SLSNe are a rare and extreme class of SN discovered in recent years \citep{2012Sci...337..927G}. The origin of SLSNe are unclear, but they play a key role in understanding the evolution of massive stars, chemical enrichment and possibly cosmic re-ionization via their bright UV luminosity. DES will discover many of these intrinsically bright objects out to redshift about 2.5 \citep{2015MNRAS.449.1215P}. At high redshifts, the optical (rest-frame UV) spectra of the SLSNe are poorly understood. Redshifts from host galaxies are thus crucial to constrain their distances and intrinsic luminosities. However, SLSNe preferentially occur in dwarf star-forming galaxies \citep{2014ApJ...787..138L}, many of which are too faint to be detected or to have reliable photo-$z$ estimates from DES multi-band imaging. OzDES provides a cost-effective inspection as the number of SLSN candidates is large with respect to the time available on 8 to 10m class telescope but is small compared to the number of AAT fibres. These SLSNe targets will gain high priority in future AAT observing seasons.

\subsection{AGN}

We obtained almost 6000 spectra of AGN and AGN candidates in Y1 (see $\S$~\ref{sec:targets} for details on the target selection strategy and Figure~\ref{fig:sample_specs} for an example spectrum).  After the completion of Y1, we identified 989 AGN to monitor. The number of AGN targets decreases from 150 per field to 100 per field, but the priority is maintained at 7 to ensure that the object is observed as frequently as possible.

\begin{figure}
\includegraphics[ width=84mm]{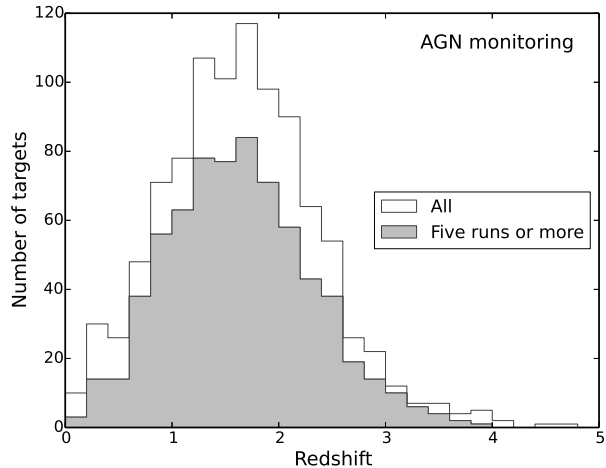}
\caption{Redshift distribution of the AGN being monitored after the first year. Shaded area represents objects that have been observed in 5 runs or more by the end of 2014.}
\label{fig:hist_rz_agn}
\end{figure}

The median redshift of these AGN is 1.63 and the distribution extends to z$\sim$4.5 (see Figure~\ref{fig:hist_rz_agn}). The sample will primarily be analyzed using either the Mg~\Rmnum{2} or C\Rmnum{4} emission line,  and in some cases both lines. We will also monitor a substantial number of targets using the $H\beta$ emission line, which is the most commonly used line in previous reverberation mapping campaigns \citep[e.g.][]{2004ApJ...613..682P}. We expect to accurately recover the radius-luminosity relationship for all three lines (King et al., in prep). 

Based on data obtained from Sept. 2013 through the end of 2014, we presently have 5 (6) or more spectroscopic epochs for 693 (455) AGN (70\% and 46\% of the sample). These numbers are close to the expected 7 epochs (in the first two seasons) for 500 AGN in our survey simulation (King et al., in prep). For more than 10\% of the sample, we have acquired spectra in 8 epochs. 

\subsection{Radio Galaxies}

By the end of Y1, we had observed 350 targets from the first data release of the ATLAS survey \citep{2006AJ....132.2409N, 2008AJ....135.1276M}, augmenting the earlier work of \citet{2012MNRAS.426.3334M}. Secure redshifts were obtained for 40\% of the targeted sources and the majority currently without redshifts are fainter than $m_r = 22.5$ mag. Sources without a redshift will be re-observed in coming campaign seasons to build up the necessary integration time to determine their redshifts and classifications. As we continue to observe these targets, we expect to increase our completeness.

With the availability of the ATLAS data release 3 (Franzen et al. in prep, Banfield et al. in prep), target selection has been modified slightly after Y1. The additional post-processing provides higher detection reliability at low radio flux levels leading to increased sensitivity to low-level star formation.

The redshifts obtained will be used to further the science goals for ATLAS including:
\begin{itemize}
\item to determine the cosmic evolution of both star forming galaxies and radio AGN, through the measurement of their redshift-dependent radio luminosity functions. For example, we will be able to measure luminosity functions for star-forming galaxies to $\sim L$* (the expected knee of the luminosity function at $z$=1).
\item to calibrate and develop photometric and statistical redshift algorithms for use with the 70 million EMU \citep{2011PASA...28..215N} sources (for which spectroscopy is impractical). 
\item to measure line widths and ratios of emission lines to distinguish star forming galaxies from AGN, low-ionization nuclear emission-line regions (LINERs), etc, and identify quasars and broad-line radio galaxies.
\item to measure how radio polarisation evolves with redshift, as a potential measure of cosmic magnetism.
\item to explore how morphology and luminosity of radio-loud AGN evolve with redshift, to understand the evolution of the jets and feedback mechanisms.
\end{itemize}

\subsection{Unusual Objects}

The large number of spectra taken by OzDES almost guarantees discoveries of rare events or objects. Even in the photometrically selected samples, outliers are expected to exist. Although an exhaustive search for unusual objects is beyond the scope of this work, we highlight this potential by noting some objects that clearly stand out when the spectra were visually inspected in \texttt{runz}. These objects fall in the following broad categories:

\begin{enumerate}

\item \textit{Unusual transients}: Time sensitive observations of live transients are of high priority. To maximize the discovery space, we target all transients that satisfy a straightforward magnitude cut. As discussed in $\S$~\ref{sec:transient_classification}, studies of transient spectra have been complicated by host galaxy light contamination and data reduction issues. A more systematic investigation will be carried out when data reduction is refined.

\item \textit{Rare stars}: At least four WD and M-dwarf binaries are recognized as Galactic objects with both a hot and a cool spectral components. We have also found a WD candidate that is likely a rare DQ WD (with carbon bands).

\item \textit{Serendipitous spectroscopic SN}: A SN may be observed unintentionally for two reasons. It happens to occur in a galaxy where a redshift is desired or it is mis-classified into a different target type. During the SV season, one of the photo-$z$ targets turned out to be a Type II SN. The chance for this to happen drops significantly in subsequent seasons as long-term photometry becomes available. However, the first possibility becomes more likely as more galaxies are targeted. Approximately one SN occurs every century in a Milky Way-like galaxy and a SN remains bright for a few weeks to a few months. It is thus expected that every few thousand galaxy spectra at relatively low redshift may contain a visible SN. The concept of spectroscopic SN search has been successfully tested \citep{2003ApJ...599L..33M,2013MNRAS.430.1746G}. However, such a survey method requires more than human eyes because the host galaxy light often dominates and has to be carefully modeled.

\item \textit{Broad Absorption Line (BAL) AGN}: A large number of AGN are observed as reverberation mapping targets, radio sources, galaxies or transients. More than a dozen of these exhibit extraordinary absorption line systems. Selected objects will be monitored throughout our survey.

\item \textit{Multiple redshifts}: If two objects fall in the same fibre, two distinct sets of spectral features may be observed. Searching for double redshifts in galaxy spectra provides a way of finding strong gravitational lens candidates \citep[e.g.][]{2004AJ....127.1860B}, particularly for small Einstein radii and faint background sources that are hard to detect by imaging. Four OzDES targets were noticed to show convincing features at two different extra-galactic redshifts. Inspection of the images reveal that two of these are merely chance alignment between background sources and foreground galaxies. The remaining two systems are lens candidates, both consisting of a foreground early type galaxy and a background emission line galaxy. 

\end{enumerate}

\section{Conclusions}\label{sec:conclusion}

OzDES is an innovative spectroscopic program that brings together the
power of multi-fibre spectrograph and time-series observations. In
five years, each DES SN survey field will be targeted in about
25 epochs during the DES observing seasons by the 2dF/AAOmega
spectrograph on the AAT. The 400 fibres are configured nightly to
target a range of objects with the goal of measuring spectroscopic
redshifts of galaxies, monitoring spectral evolution of AGN or
classifying transients. Stacking of multi-epoch spectra allows OzDES
to measure redshifts for galaxies that are as faint as $m_r=25$ mag. Along
with efficient redshifting and recycling of fibres, we expect to
obtain about 2,500 host galaxy redshifts for the DES SN cosmology
study. The long term time series, contemporaneous with DECam imaging,
will enable reverberation mapping of the largest AGN sample to
date. In addition, OzDES is an important source of redshifts for
various DES photo-$z$ programs.

In the above sections, we have summarized the OzDES observing strategy, reported results from the first year of operation and evaluated the outcome in light of various science goals. Overall, our strategy has worked well in the first year and produced a large number of redshifts with good quality. We are on track to achieve our main science goals. Meanwhile, we have identified a number of areas for improvements, including better data reduction procedures to reduce artefacts and more rigorous cross-check to raise redshift reliability. 

By the time of writing, the second observing season (Y2) has completed and already a number of updates have been implemented. Additional data and better understanding of the survey yield has allowed us to reassess target selection and de-selection criteria for different target types and science goals. For example, targeting the brightest objects first maximizes the number of measured redshifts for radio galaxies, but such tactics only work when selection bias is not a major consideration. It is also desirable to abandon an unsuccessful target after certain number of exposures while the maximum integration time allowed depends on the total number of redshifts anticipated for this particular target type. Short integration times allow faster recycling of fibres and more redshifts to be measured. During poor observing conditions, a backup program is in place to measure redshifts for bright galaxies, along with high priority targets such as AGN and active transients. 

As for all long-term projects, we expect to continue to refine our data quality, actively analyze new data and adapt the observing strategy in the coming years. 

\section*{Acknowledgments}

Parts of this research were conducted by the Australian Research Council Centre of Excellence for All-sky Astrophysics (CAASTRO), through project number CE110001020.
ACR acknowledges financial support provided by the PAPDRJ CAPES/FAPERJ Fellowship.
FS acknowledges financial support provided by CAPES under contract No. 3171-13-2
BPS acknowledges support from the Australian Research Council Laureate Fellowship Grant LF0992131.
This work was supported in part by the U.S. Department of Energy contract to SLAC No. DE-AC02-76SF00515.

The data in this paper were based on observations obtained at the Australian Astronomical Observatory (AAO programs A/2012B/11 and A/2013B/12, and NOAO program NOAO/0278). We'd like to thank Marguerite Pierre and the XMM-XXL collaboration for allowing us to use a couple of hours of their time on the AAT to target the DES C3 field.

The authors would like to thank Charles Baltay and the La Silla Quest Supernova Survey to conduct a concurrent transient search during the SV season to help test the targeting strategy.
We are grateful for the extraordinary contributions of our CTIO colleagues and the DECam Construction, Commissioning and Science Verification
teams in achieving the excellent instrument and telescope conditions that have made this work possible.  The success of this project also 
relies critically on the expertise and dedication of the DES Data Management group.

Funding for the DES Projects has been provided by the U.S. Department of Energy, the U.S. National Science Foundation, the Ministry of Science and Education of Spain, 
the Science and Technology Facilities Council of the United Kingdom, the Higher Education Funding Council for England, the National Center for Supercomputing 
Applications at the University of Illinois at Urbana-Champaign, the Kavli Institute of Cosmological Physics at the University of Chicago, 
the Center for Cosmology and Astro-Particle Physics at the Ohio State University,
the Mitchell Institute for Fundamental Physics and Astronomy at Texas A\&M University, Financiadora de Estudos e Projetos, 
Funda{\c c}{\~a}o Carlos Chagas Filho de Amparo {\`a} Pesquisa do Estado do Rio de Janeiro, Conselho Nacional de Desenvolvimento Cient{\'i}fico e Tecnol{\'o}gico and 
the Minist{\'e}rio da Ci{\^e}ncia e Tecnologia, the Deutsche Forschungsgemeinschaft and the Collaborating Institutions in the Dark Energy Survey. 

The DES data management system is supported by the National Science Foundation under Grant Number AST-1138766.
The DES participants from Spanish institutions are partially supported by MINECO under grants AYA2012-39559, ESP2013-48274, FPA2013-47986, and Centro de Excelencia Severo Ochoa SEV-2012-0234, 
some of which include ERDF funds from the European Union.

The Collaborating Institutions are Argonne National Laboratory, the University of California at Santa Cruz, the University of Cambridge, Centro de Investigaciones Energeticas, 
Medioambientales y Tecnologicas-Madrid, the University of Chicago, University College London, the DES-Brazil Consortium, the Eidgen{\"o}ssische Technische Hochschule (ETH) Z{\"u}rich, 
Fermi National Accelerator Laboratory, the University of Edinburgh, the University of Illinois at Urbana-Champaign, the Institut de Ciencies de l'Espai (IEEC/CSIC), 
the Institut de Fisica d'Altes Energies, Lawrence Berkeley National Laboratory, the Ludwig-Maximilians Universit{\"a}t and the associated Excellence Cluster Universe, 
the University of Michigan, the National Optical Astronomy Observatory, the University of Nottingham, The Ohio State University, the University of Pennsylvania, the University of Portsmouth, 
SLAC National Accelerator Laboratory, Stanford University, the University of Sussex, and Texas A\&M University.

This paper has gone through internal review by the DES collaboration.

\section*{Affiliations}
{\small
$^{1}$\RSAA \\ 
$^{2}$\CAASTRO \\ 
$^{3}$\AAO \\ 
$^{4}$\UQ \\ 
$^{5}$\UCL \\ 
$^{6}$\KavliCam \\ 
$^{7}$\Cambridge \\ 
$^{8}$\FNAL \\ 
$^{9}$\ON \\ 
$^{10}$\LIneA \\ 
$^{11}$\NDphy \\ 
$^{12}$\INAF \\ 
$^{13}$\IEEC \\ 
$^{14}$\Portsmouth \\ 
$^{15}$\KavliStanford \\ 
$^{16}$\UIUCastro \\ 
$^{17}$\UIUCphy \\ 
$^{18}$\KavliChicago \\ 
$^{19}$\Swinburne \\ 
$^{20}$\LBNL \\ 
$^{21}$\Dark \\ 
$^{22}$\ANL \\ 
$^{23}$\USyd \\ 
$^{24}$\OSUccapp \\ 
$^{25}$\OSUastro \\ 
$^{26}$\CSIRO \\ 
$^{27}$\CAPES \\ 
$^{28}$\Sussex \\ 
$^{29}$\SLAC \\ 
$^{30}$\UArizona \\ 
$^{31}$\UP \\ 
$^{32}$\Southampton \\ 
$^{33}$\Stanfordphy \\ 
$^{34}$\CTIO \\ 
$^{35}$\IAP \\ 
$^{36}$\NCSA \\ 
$^{37}$\TAMU \\ 
$^{38}$\LMU \\ 
$^{39}$\JPL \\ 
$^{40}$\UMphy \\ 
$^{41}$\MPI \\ 
$^{42}$\Munich \\ 
$^{43}$\OSUphy \\ 
$^{44}$\ICRA \\ 
$^{45}$\UMastro \\ 
$^{46}$\IFAE \\ 
$^{47}$\ICREA \\ 
$^{48}$\BNL \\ 
$^{49}$\CIEMAT \\ 
}

\end{document}